\documentclass{article}
\usepackage{iclr2027_conference,times}
\usepackage{microtype}
\usepackage{graphicx}
\usepackage{float}
\usepackage{booktabs}
\usepackage{longtable}
\usepackage{array}
\usepackage{enumitem}
\usepackage{xcolor}
\usepackage{amsmath,amssymb}
\usepackage{url}
\usepackage[labelsep=period]{caption}
\usepackage[hidelinks]{hyperref}
\hypersetup{
  pdftitle={LiveUIBench: Evaluating Agent as App Workflows},
  pdfauthor={Tong Zhang, Yuxuan Cheng, Ruimin Zhang, Yun Peng, Tao Xie}
}
\title{LiveUIBench: Evaluating \\Agent as App Workflows}
\author{Tong Zhang\textsuperscript{1} \quad Yuxuan Cheng\textsuperscript{2} \quad Ruimin Zhang\textsuperscript{3} \quad Yun Peng\textsuperscript{1} \quad Tao Xie\textsuperscript{4}\\
{\normalfont\textsuperscript{1}Fudan University \quad \textsuperscript{2}The Chinese University of Hong Kong, Shenzhen}\\
{\normalfont\textsuperscript{3}Tongji University \quad \textsuperscript{4}Peking University}}
\newcommand{\bench}{\textsc{LiveUIBench}}
\newcommand{\liveuiagent}{\textsc{LiveUIAgent}}

\newcommand{\vendorlogo}[1]{\raisebox{-0.12ex}{\includegraphics[height=0.9em,width=1.0em,keepaspectratio]{revisions/yun/figures/vendor/#1.pdf}}\hspace{0.25em}}
\newif\ifvfourappendix
\ifdefined\VFOURMAINONLY\vfourappendixfalse\else\vfourappendixtrue\fi
\iclrfinalcopy
\begin{document}
\raggedbottom
\maketitle
\lhead{}
\begin{abstract}
Applications provide convenient features for users to complete daily tasks.
However, different users may have different or even conflicting requirements
for the same application, while the requirements of the same user may change
over time.
Personalizing applications produced by agent-to-app or UI-to-code approaches
can require changes to the generated implementation.
We propose Agent as App, a task in which users express their requirements
in natural language and an agent creates a personalized application while
continuing to supply its behavior during use.
We implement \liveuiagent{}, which generates a task-specific orchestration
connecting the user interface, saved information, and service actions.
A shared agent runtime renders and updates the interface while carrying out
the requested operations.
Revising the orchestration and interface while reusing that runtime provides
a lightweight basis for personalization as requirements evolve.
We introduce \bench{}, a benchmark with diverse requirements and ordered
interactions across personal services.
The empirical findings show that the evaluated models struggle to complete
these workflows reliably.
Applications can satisfy many local checks while leaving the overall task
unfinished, and greater partial coverage can coincide with fewer completed
workflows.
\end{abstract}
\section{Introduction}
Applications help users complete everyday activities, but people can have
different or conflicting requirements for the same application.
A music listener may want a simple play button, while another wants to
compare candidates and maintain a listening journal.
Personalization also changes over time. A listener can begin with a
playlist and later ask to keep notes about songs, revise earlier choices,
or prepare a selection for an event.
The application must connect each request to information already saved.
Generative and malleable interfaces make such variation a concrete design
goal~\citep{dynavis2024,malleableoverview2025,malleableui2025}.
Adaptive and mixed-initiative interfaces provide earlier foundations for
matching interaction to a user's needs~\citep{supple2004,horvitz1999}.

Language agents provide a way to express these requirements directly.
An agent can produce an application implementation, a producer role we call
agent-to-app.
End-user software and data-driven authoring systems connect user intent
to executable behavior~\citep{ko2011,mavo2016}.
UI-to-code systems translate a visual design into an
implementation~\citep{design2code2025}, while website-generation benchmarks
test whether produced artifacts support their requested
functions~\citep{webgenbench2025}.
An agent can also remain responsible for interpreting task actions after
an interface has been created.
The application then combines visible interface elements, continued agent reasoning,
and saved information. Its behavior depends on their connection during use.

\textbf{We propose Agent as App as a task for persistent, personalized
applications.}
Given a natural-language application requirement and permitted context,
an agent must create an interactive workspace and continue to provide its
task behavior as the user supplies inputs, changes requests, and returns.
The application-side agent interprets inputs and retained information,
performs service operations, and updates the interface.
Success concerns the requested effects across this sequence.
Presentation-only events can be handled without invoking a model.
The task leaves the orchestration format and interface layout open.

\begin{figure}[t]
\centering
\includegraphics[width=\linewidth]{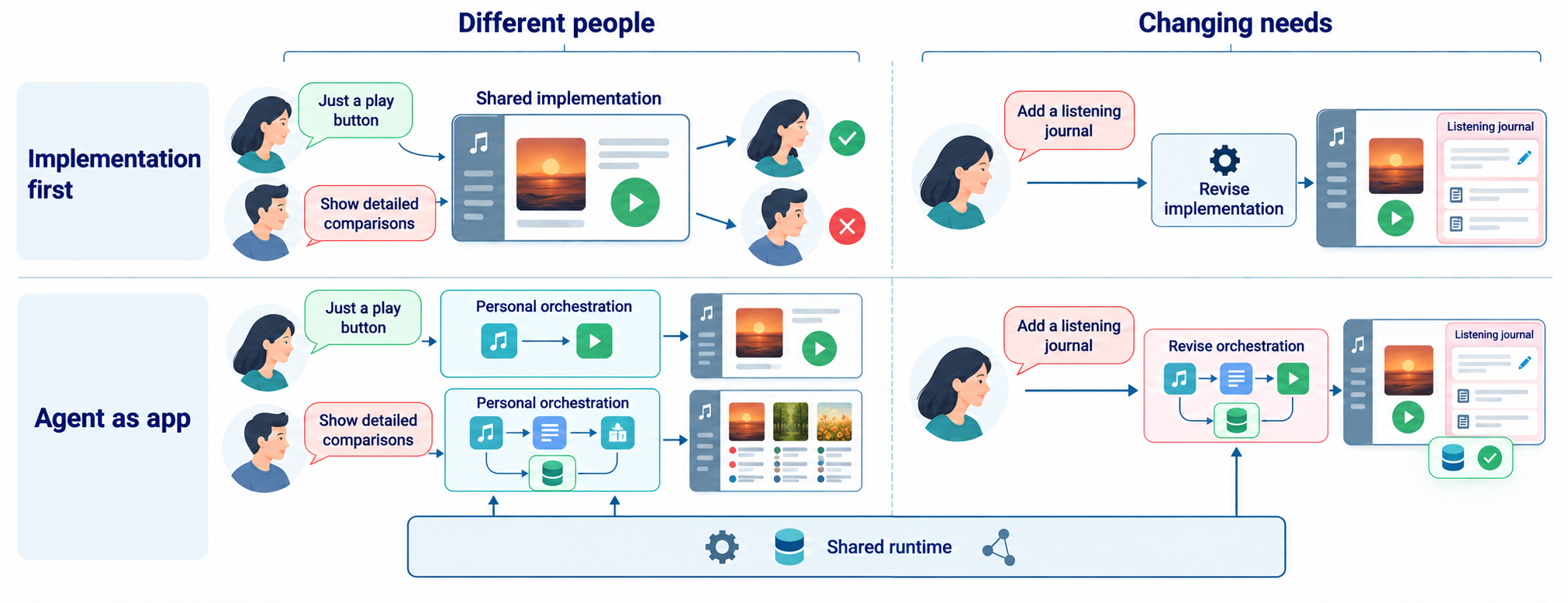}
\caption{Agent as App connects personalization with continued use.
Different listeners request a minimal player or detailed comparisons.
A later request adds a listening journal that retains information alongside
the original activity. The generated application must connect these goals
to its interface, behavior, and saved information.}
\label{fig:story}
\end{figure}

The evaluation object is the application through continued interaction.
A convincing initial screen does not establish that its interface reaches the
intended service operations. A successful service call can affect the wrong
record. A return visit can preserve a mistake.
These outcomes look similar if evaluation records only whether a screen opens,
an action returns, or stored information remains present.
They differ when the requested goal, visible result, and saved effect are
checked together.

We implement Agent as App in \liveuiagent{}.
A user's requirement produces an orchestration that links the generated
interface to service actions and saved application state.
The agent remains responsible for executing later requests through this
interface. Revising a requirement updates the orchestration and workspace,
while a shared runtime continues to execute actions and retain the user's
information. Personalization can therefore proceed through changes to
the task-specific plan without rebuilding the execution machinery.

\bench{} evaluates these applications through sequences of user requests.
An initial requirement is followed by submissions, revisions, and return
visits across personal services. The benchmark measures whether the
generated application completes the sequence, alongside its coverage of
individual assertions. We analyze the underlying task records to locate
the earliest unmet requirements and compare which tasks different models
complete. The results reveal a large gap between opening an application
and completing its workflow. They also show that stronger partial
assertion coverage can coexist with fewer completed tasks, and that
combining the successes of different models still leaves much of the
benchmark unresolved.

Our contributions are listed below.
\begin{enumerate}[leftmargin=*,itemsep=2pt,topsep=3pt]
\item \textbf{We propose the Agent as App task.}
The task connects personalized application creation with continued
agent-mediated use as users supply inputs, revise requirements, and return.
\item \textbf{We implement \liveuiagent{}.}
The implementation connects personalized interfaces to a shared agent
runtime and saved service state through lightweight orchestrations.
\item \textbf{A benchmark and empirical findings.}
\bench{} provides diverse task requirements and ordered interactions.
The empirical analysis follows failure stages, unmet checks, and the
shared and distinct tasks completed by different models.
\end{enumerate}

\section{LiveUIBench}
\label{sec:task}
\paragraph{Requirements across users and interactions.}
The requirements of user $u$ at interaction stage $t$ are represented by
$r_{u,t}$.
Different users can ask for different behavior at the same stage,
so $r_{u,t}\neq r_{v,t}$.
A user can also revise a requirement, so $r_{u,t+1}\neq r_{u,t}$.
Here $t$ orders interactions. It does not measure elapsed calendar time.

These changes need not be large.
A restocking application can begin by recording household items and later
be asked to revise a selected item without replacing the rest of the list.
A music application can retain candidates while the user changes a decision.
Personalization therefore includes interpreting a new request in relation
to what the application already knows and has changed.
The task connects this interpretation to visible interface elements and saved effects.

\paragraph{Application lifecycle.}
An application is represented by $A_{u,t}=(P_{u,t},S_{u,t})$.
Here $P$ specifies its interface and behavior, and $S$ is its saved state.
The service world is $W_{u,t}$, and an input submitted through the application
is $x_{u,t}$. Creation and continued use take the form
\begin{equation}
\begin{aligned}
A_{u,0} &= G(r_{u,0},c_u),\\
(A_{u,t+1},W_{u,t+1},V_{u,t+1})
  &= F(A_{u,t},W_{u,t},x_{u,t+1}),
\end{aligned}
\label{eq:agent-as-app-task}
\end{equation}
where $c_u$ is the context available at creation and $V$ is the resulting
visible interface.
Later operations use the created application and its records.
An ordinary input can leave $r$ unchanged. A revision communicated through
$x$ asks for changed behavior or saved values while retaining relevant information.
A return visit checks whether that information remains available.

\paragraph{Tasks and episodes.}
A benchmark task supplies an initial requirement, starting context, and
ordered interaction goals.
A task evaluation applies it to a model or condition with a retained
generated application.
An episode supplies a user goal and predefined assertions about actions or state.
Episodes cover opening, submission, revision, and return, with follow-up
requests wherever the task specifies them.
Their order matters. A revision refers to a result created earlier, and
a later application may need to discover a record written by another application.

\paragraph{Workflow success.}
At stage $t$, the requested goal is $g_{u,t}$ and the service operations are $L_{u,t}$.
The task defines a predicate $Q_{u,t}$ over the visible result, application
state, service state, and operation trace.
Complete workflow success is
\begin{equation}
y_u=\prod_{t=1}^{T_u}
\mathbf{1}\!\left[
Q_{u,t}(g_{u,t},V_{u,t},S_{u,t},W_{u,t},L_{u,t})
\right].
\label{eq:workflow-success}
\end{equation}
Each predicate encodes the episode goal through its predefined assertions.
A local success cannot substitute for a missing requirement elsewhere in
the sequence. The evaluation checks which record an action affects, how
that record changes, what the interface shows, and what remains on return.

\paragraph{Task organization.}
\bench{} covers Spotify, Amazon, Gmail, Phone, Venmo, Splitwise, Todoist,
Simple Note, and File System.
AppWorld supplies the executable service environment~\citep{appworld2024}.
Single-application tasks exercise an ordered workflow in a retained workspace.
Multiple-tool tasks coordinate service operations within an application.
Shared-world tasks require a later application to use records written earlier.
These families describe the organization of the workflow.
The service dependencies are counted separately in Table~\ref{tab:task-metrics}.

\begin{table}[t]
\caption{Task coverage and service dependencies. Family labels describe application organization. Cross-service counts are independently computed from distinct named services. Counts precede repeated generation draws.}
\label{tab:task-metrics}
\centering\small
\setlength{\tabcolsep}{3pt}
\renewcommand{\arraystretch}{1.08}
\begin{tabular}{@{}>{\raggedright\arraybackslash}p{\dimexpr.20\linewidth-4pt}>{\raggedleft\arraybackslash}p{\dimexpr.15\linewidth-4pt}>{\raggedleft\arraybackslash}p{\dimexpr.21\linewidth-4pt}>{\raggedright\arraybackslash}p{\dimexpr.44\linewidth-6pt}@{}}
\toprule
\textbf{Task family} & \textbf{All/public} & \textbf{Cross-service} & \textbf{Dependency} \\
\midrule
Single application & 302/93 & 51/9 & One orchestration handles ordered goals. \\
Multiple tools & 130/73 & 130/73 & One orchestration coordinates three or four services. \\
Shared world & 68/6 & 58/5 & A later application uses service state written by an earlier application. \\
Service tasks & 27 extra & 0 & One application creates, revises, and reads a record, then reopens. \\
\bottomrule
\end{tabular}
\par\vspace{4pt}
\begin{minipage}{\linewidth}\small
The resource contains 500 tasks, 2,408 episodes, and 10,590 assertions,
including the episode groups of shared-world tasks.
Its development, public, and exploratory partitions contain 147, 172, and 181 tasks.
Family labels describe application organization. Service dependencies are counted separately.
Additional service tasks support record-centered evaluation.
\end{minipage}
\end{table}

\paragraph{Record-centered service tasks.}
The additional service tasks follow a specified entity through creation,
revision, retrieval, and return.
Independent reads use the entity's stable identifier to check its requested fields.
This makes a service effect concrete.
The evaluator can distinguish a changed priority from an acknowledgment,
an edit to the original record from creation of a replacement, and current
information from an earlier stored view.
The recorded examples in Appendix~\ref{app:case-studies} connect these checks
to the interface through which a user requests a change.

\section{Method}
\label{sec:system}
\begin{figure}[!t]
\centering
\includegraphics[width=\linewidth]{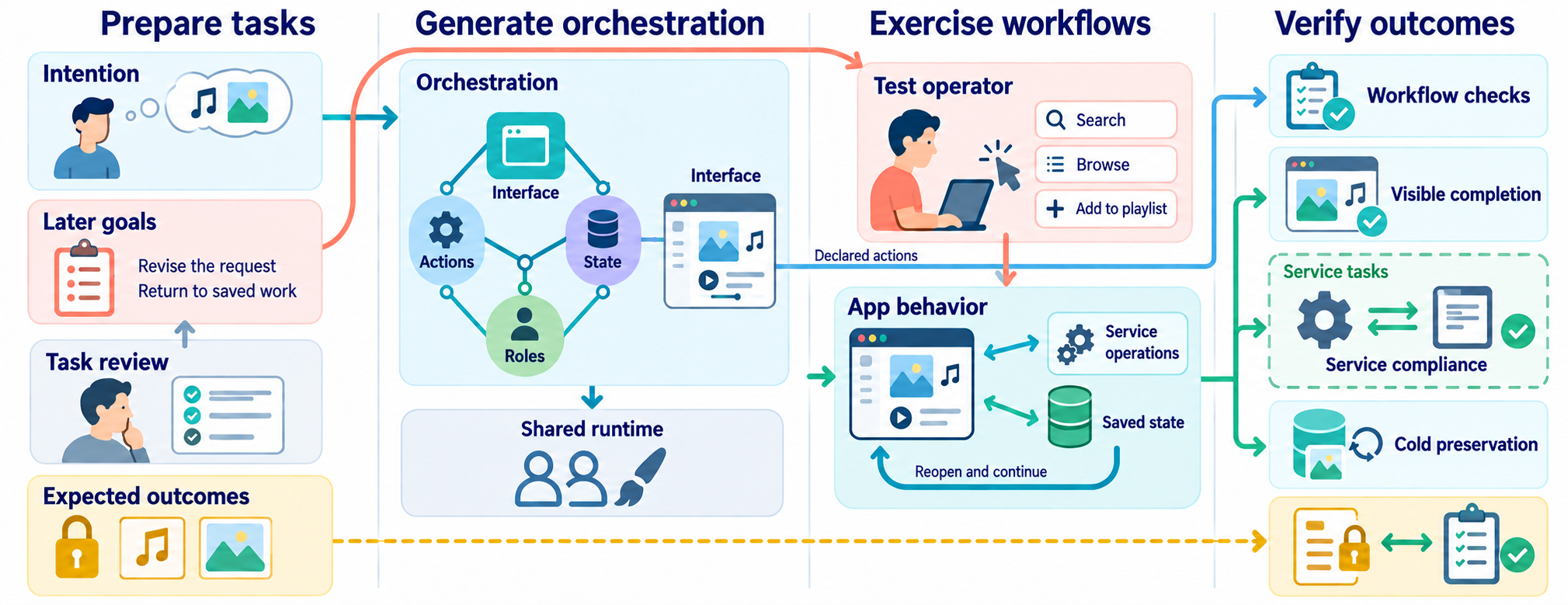}
\caption{The evaluation workflow.
Prepare an intention, later goals, and expected outcomes.
Generate an orchestration from the intention and permitted context.
Execute ordered goals through declared actions or the rendered interface.
Check workflow completion, visible outcomes, and preserved state.
Service tasks additionally verify saved records independently.
The test operator supplies user actions. The application-side agent
interprets them and carries out the requested operations.}
\label{fig:evaluation}
\end{figure}

Figure~\ref{fig:evaluation} follows application creation, continued interaction,
and verification of the requested outcomes.

\subsection{Prepare the task}
A task supplies an initial requirement, service context, and ordered
episode requests.
The starting service world contains the records assumed by that task.
The generator receives the initial requirement and permitted service
documentation. Later requests arrive at the corresponding episode.
Task specifications and assertions are checked for agreement with the
user's goal, as detailed in Appendix~\ref{app:human-review}.

\subsection{Generate the application}
\liveuiagent{} generates an orchestration linking interface elements to action
goals, agent roles, and information retained across interactions.
An action goal describes what a submitted action should achieve and
which services it may use~\citep{react2023,apibank2023}.
A UI designer produces the component layout.
The shared renderer presents that layout using current inputs and saved state.

\begin{figure}[t]
\centering
\includegraphics[width=\linewidth]{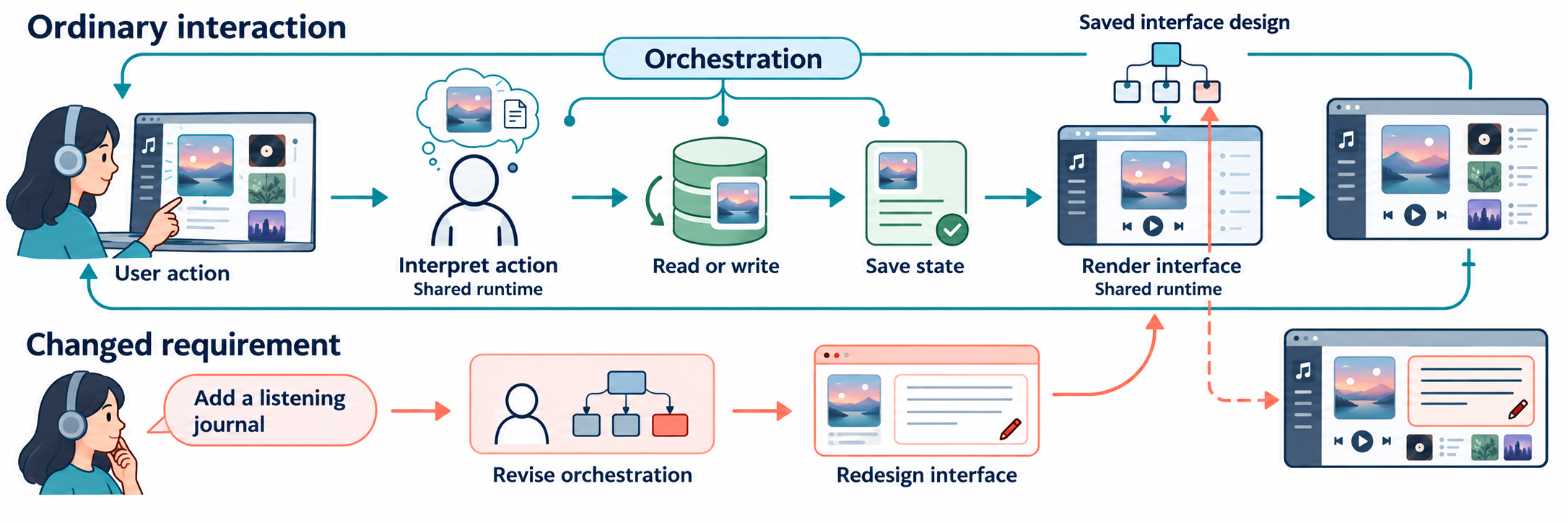}
\caption{A task-specific orchestration links interface elements, action goals, and saved information.
The shared runtime interprets submitted inputs, uses services, saves the
result, and renders the resulting interface. The application keeps its
identity as the user continues the workflow.}
\label{fig:runtime}
\end{figure}

The application-side agent interprets submitted inputs, performs service
operations, and saves a result for the interface to present.
An ordinary action uses the existing orchestration.
A changed requirement can instead revise the orchestration and generate
an updated interface while retaining the application's identity and information.
Appendix~\ref{app:guios-workspace} shows the complete workspace.

\subsection{Run the application}
The task operator supplies the current episode's inputs, follows revisions
and extensions, and performs the specified return visit.
The application-side agent executes the submitted request using its current
inputs, saved state, and available service responses.
The model survey invokes declared actions with the episode requests.
The recorded application examples additionally follow the visible interface.

The evidence retains submitted inputs, application responses, and service
operations in their original order.
Later episodes can refer to information established earlier, including
information produced by an application that shares the same service world.
A return visit reads the retained result instead of recreating it.

\subsection{Check the result}
The requested goal determines which record or state change matters.
Visible evidence describes the result presented to the user.
Service observations establish the recorded operations and saved effects.
For record-centered examples, independent reads follow the same entity
through creation, revision, retrieval, and return.

A priority revision requires the requested value on the intended record.
An acknowledgment alone establishes only that the interface responded.
Likewise, recovery after reopening establishes persistence of the retained
information, whose correctness must still be checked against the request.
The examples in Appendix~\ref{app:case-studies} examine these distinctions
through the same application and record across successive interactions.

\section{Evaluation}
\label{sec:evaluation}
The model survey evaluates generated orchestrations against the benchmark's
original operational assertions.
Opening checks test whether the expected views, inputs, and action bindings
are available. Later assertions examine service calls, changes to saved
information, evidence in the displayed result, and retained state.
Table~\ref{tab:multimodel} reports the model configurations and task-family results.

The full-workflow rate is the fraction of tasks satisfying every assigned
assertion. The assertion pass rate measures coverage across the complete
set of assigned assertions.
These measures establish the recorded checks; semantic requirements beyond
those checks need separate service evidence.
Opening coverage requires every opening episode within a task to pass,
including the later application in a shared-world task.
The metrics retain rejected generations and unavailable responses in
their assigned denominators. The survey uses the recorded generation
procedures for each model, so comparisons describe the evaluated systems.
Metric definitions and the survey display rule appear in
Appendix~\ref{app:model-survey}.

We recompute the measures from the archived per-task and per-episode records.
For each failed task, the first episode containing an unmet assertion
determines its failure stage.
The failed checks in that episode identify what evidence was missing.
Recorded generation and episode errors remain separate from completed
operations. This analysis locates observed failures within the prescribed
sequence without inferring an internal cause from the stage label.

We also compare the sets of tasks completed by different models.
Their union measures which tasks have been completed by at least one
evaluated model. It is a retrospective coverage diagnostic, not a measured
ensemble or a model-selection procedure.
The recorded application examples provide a complementary view of
interaction and saved service effects.
Appendix~\ref{app:raw-analysis} specifies the recount and grouping rules.

\section{Results}
\label{sec:experiments}
Table~\ref{tab:multimodel} summarizes model performance.
We then examine the underlying task records to determine where workflows
fail, which checks remain unmet, and whether different models solve the
same tasks.

\begin{table}[!ht]
\caption{Model results on 500 assigned tasks.
Open counts tasks passing every opening check. Single, Multi-tool, Shared,
and Overall report FWR (\%). APR covers 10,590 assertions (\%).
Models are grouped by developer and ordered by FWR within each group.
Bold marks the highest result in each column.}
\label{tab:multimodel}
\centering\small
\setlength{\tabcolsep}{3pt}
\renewcommand{\arraystretch}{1.08}
\begin{tabular*}{\linewidth}{@{\extracolsep{\fill}}llrrrrrr@{}}
\toprule
\textbf{Developer} & \textbf{Model} & \textbf{Open} & \shortstack{\textbf{Single}\\$N=302$} & \shortstack{\textbf{Multi-tool}\\$N=130$} & \shortstack{\textbf{Shared}\\$N=68$} & \textbf{Overall} & \textbf{APR} \\
\midrule
Amazon & \vendorlogo{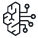} Nova Premier & 152 & 2.98 & 1.54 & 0.00 & 2.20 & 22.74 \\
\addlinespace[2pt]
DeepSeek & \vendorlogo{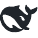} DeepSeek V4 Pro & 317 & 9.93 & \textbf{5.38} & 0.00 & 7.40 & 44.00 \\
\addlinespace[2pt]
Google & \vendorlogo{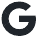} Gemini 3.8 Flash & 463 & 17.55 & 3.08 & 7.35 & 12.40 & 65.07 \\
\addlinespace[2pt]
Meta & \vendorlogo{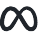} Llama 4 Maverick & 27 & 0.33 & 0.00 & 0.00 & 0.20 & 4.19 \\
\addlinespace[2pt]
MiniMax & \vendorlogo{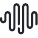} MiniMax M3 & 33 & 0.66 & 0.00 & 0.00 & 0.40 & 4.62 \\
\addlinespace[2pt]
Mistral & \vendorlogo{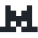} Mistral Medium 3.5 & 105 & 1.32 & 0.77 & 0.00 & 1.00 & 14.83 \\
\addlinespace[2pt]
Moonshot & \vendorlogo{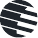} Kimi K3 & 395 & 20.53 & 4.62 & 5.88 & 14.40 & 61.11 \\
\addlinespace[2pt]
NVIDIA & \vendorlogo{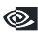} Nemotron 3 Ultra & 175 & 8.61 & 1.54 & 0.00 & 5.60 & 24.74 \\
\addlinespace[2pt]
OpenAI & \vendorlogo{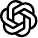} GPT-5.6-sol & 483 & 16.23 & 3.85 & 4.41 & 11.40 & 71.09 \\
 & \vendorlogo{openai} GPT-5.6-luna & 331 & 11.59 & 3.85 & 2.94 & 8.40 & 48.59 \\
\addlinespace[2pt]
Tencent & \vendorlogo{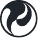} Hunyuan 3 & 175 & 6.62 & 2.31 & 0.00 & 4.60 & 26.33 \\
 & \vendorlogo{hunyuan} Hunyuan 4 Preview & 108 & 3.97 & 0.77 & 0.00 & 2.60 & 17.53 \\
\addlinespace[2pt]
xAI & \vendorlogo{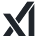} \textbf{Grok 4.6} & 491 & \textbf{28.48} & 3.85 & \textbf{10.29} & \textbf{19.60} & \textbf{76.61} \\
 & \vendorlogo{xai} Grok 4.5 & \textbf{496} & 22.19 & 3.08 & 7.35 & 15.20 & 73.42 \\
\addlinespace[2pt]
Z.ai & \vendorlogo{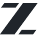} GLM-5.2 & 329 & 13.25 & 4.62 & 2.94 & 9.60 & 50.04 \\
\bottomrule
\end{tabular*}
\end{table}

\begin{figure}[!t]
\centering
\includegraphics[width=\linewidth]{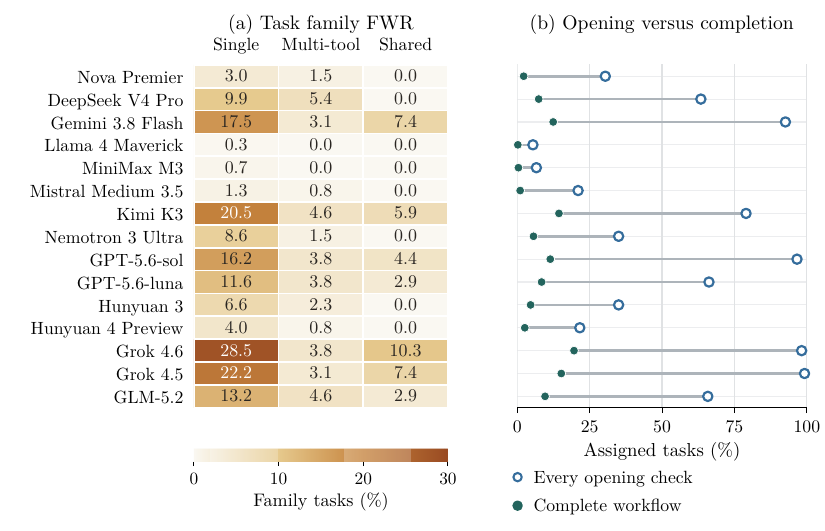}
\caption{Task-family performance and the opening-to-completion gap.
All 15 models from Table~\ref{tab:multimodel} share the same row order.
(a) Each cell reports full-workflow success within its task family, using
302 single-application, 130 multi-tool, and 68 shared-world tasks.
The color scale is shared across all cells.
(b) Opening and full-workflow rates use the same 500 assigned tasks per model.
The connector shows the gap between the two task-level measures.}
\label{fig:survey-profiles}
\end{figure}

\paragraph{Opening an application is much easier than completing its workflow.}
The leading models pass opening checks on nearly the entire task collection,
yet complete workflows remain uncommon.
Figure~\ref{fig:survey-profiles} compares task-family completion with the gap between opening and
completing a workflow across all displayed models.
An application can expose the requested inputs and actions while a later
submission, revision, or return fails.
This gap remains among models with high assertion coverage, so broad
partial progress gives an incomplete picture of the applications a user
can rely on through an entire task.

Both endpoints use the same assigned tasks, so the gap can be interpreted
within each model.
Passing all opening checks establishes that the prescribed interfaces and
entry points are available. Completing the workflow additionally requires
the later actions to satisfy their assigned checks in those applications.
The long connectors in Figure~\ref{fig:survey-profiles} show how frequently
these requirements separate, even when application opening is rarely the
earliest source of failure.

\paragraph{The ranking depends on what must succeed together.}
Sol has higher opening and assertion coverage than Kimi, while Kimi
completes more workflows.
Kimi's workflow advantage holds within every reported task family.
Thus, the reversal persists when the task collection is separated by its
workflow organization.
A model can satisfy more individual checks without concentrating those
successes on applications that meet all of a user's requirements.

A similar reversal appears between the Grok variants.
The variant with broader opening coverage completes fewer workflows,
while the stronger workflow variant also has higher assertion coverage.
Opening readiness therefore does not consistently preserve the ranking
obtained from either later measure.

The family columns also change the overall model ordering.
DeepSeek leads the multi-tool family despite placing below several models
in overall completion.
Its advantage does not extend to shared-world tasks, where it completes
no workflow. Grok leads overall and in the single-application and
shared-world families.
These differences describe the evaluated systems and task collection.
They do not isolate the effect of task structure from the differing
requirements and generation histories.

\begin{figure}[!t]
\centering
\includegraphics[width=\linewidth]{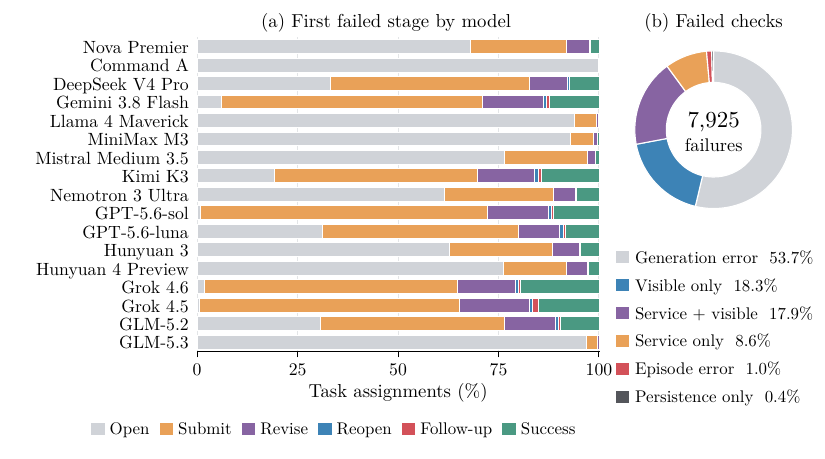}
\caption{Failure location and recorded evidence in the original survey.
(a) Each model's 500 tasks are partitioned by the first failed native
episode or success. All 17 selected models are included.
Open can include a later application in a shared-world task.
(b) The 7,925 failed task assignments are grouped by the failed checks in
their first failed episode, with generation and episode errors separate.
Service and visible categories can overlap within an episode, and the
combined category keeps the displayed groups mutually exclusive.
These are recorded failure patterns, not diagnoses of internal causes.}
\label{fig:survey-failures}
\end{figure}

\paragraph{Submission and revision account for the main loss after opening.}
Figure~\ref{fig:survey-failures} assigns each task to its first failed
episode or to success.
Models with low opening coverage frequently fail during application
generation. For the strongest workflow models, submitting a request is the dominant
first failure. Revision is another common first-failure stage after a passing submission.
The difficult transition is from an available interface to an application
that carries out the requested operation.

Reopening accounts for a small share of first failures.
This does not establish reliable persistence, because a task that has
already failed during submission is assigned to that earlier stage.
The first-failure analysis localizes the earliest unmet requirement,
while later outcomes remain in the task record.
Opening stages can also recur in shared-world tasks, so an opening failure
need not concern the first application in the sequence.

\paragraph{Service activity and a supported visible result can disagree.}
The failure profiles separate missing service evidence from missing
support in the displayed result.
Some episodes pass their recorded service checks but fail the visible
checks, while others fail both.
Appendix~\ref{app:native-shift-case} follows a recorded application
through these distinct checks.
The service checks concern recorded calls and changes, rather than every
semantic detail of the saved values.

The episode-level comparison provides a complementary view of this gap.
It includes episodes in which the visible checks pass while service
checks fail, as well as episodes with the opposite outcome.
The first pattern leaves the displayed result without all of the required
service evidence. The second records service activity without all of the
required support in the visible result.
A complete task must join these requirements across the same sequence.
The first-failure figure counts each failed task once, while the joint
check analysis in Appendix~\ref{app:raw-analysis} retains the individual
episodes on which these disagreements occur.

The disagreement is asymmetric: passing service checks while failing
visible checks occurs more often than the reverse among the episodes
with both kinds of evidence.
Service activity alone is consequently an incomplete account of the
observed failures, even when the required calls or changes are recorded.
The visible checks add requirements that those service checks do not
capture. This distinction matters when interpreting partial progress:
a recorded operation can contribute to assertion coverage while the
application still fails to present a supported result to its user.

\paragraph{Different successes leave a large common set of unsolved tasks.}
Comparing task identities reveals complementary successes that aggregate
scores conceal.
Combining the observed successful tasks of all evaluated models expands
coverage beyond the strongest model, but most tasks remain unresolved by
every model.
The shared difficulty is particularly clear in the multi-tool and
shared-world families, where even this union covers few tasks.
The union is a retrospective diagnostic of existing results, rather than
a tested ensemble. Table~\ref{tab:survey-coverage} reports the counts and
task-family denominators.

The success sets are also not nested in the aggregate ranking.
A model with fewer completed tasks can still complete tasks that a
higher-scoring model fails.
This explains why pooling observed successes increases coverage, but
the gain does not remove the large unresolved part of the collection.
The union is useful for separating shared difficulty from complementary
strengths. It does not establish that the successful model can be
identified before executing a task, and it does not count a partially
successful application as a completed workflow.

\section{Related Work}
\label{sec:related-work}
End-user software engineering studies how people develop software for their
own tasks~\citep{ko2011}, and Mavo connects authored interfaces to persistent
user data~\citep{mavo2016}.
Mixed-initiative interfaces and SUPPLE adapt interaction to users and
devices~\citep{horvitz1999,supple2004}.
Recent generative and malleable interfaces extend this flexibility through
language and dynamically produced
views~\citep{generativeinterfaces2025,dynavis2024,malleableoverview2025,malleableui2025}.
Agent as App follows the application beyond its initial presentation.
The generated interface must support later requests, preserve relevant
information, and connect a user's changing intention to service behavior.

Reusable automation offers a related way to turn user intentions into
repeatable actions.
Sikuli, SUGILITE, PUMICE, and CoScripter use visual references, demonstrations,
or accessible scripts to specify
behavior~\citep{sikuli2009,sugilite2017,pumice2019,coscripter2008}.
Language agents combine reasoning with external
actions~\citep{react2023,apibank2023,toolbench2023}, while workflow memory
supplies reusable execution structure~\citep{agentworkflowmemory2024}.
Self-Refine and Reflexion improve outputs or later attempts through
feedback~\citep{selfrefine2023,reflexion2023}, and generative agents maintain
memory for continuing behavior~\citep{generativeagents2023}.
Agent as App follows requests made within a retained application,
where earlier information and service effects constrain each subsequent action.

Feedback during an attempted solution and a new request from the user
impose different requirements on an application.
A new request can revise the intended outcome while preserving earlier
information that remains relevant.
For example, extending a gift queue should retain the entries the user
still needs, while revising a task should preserve the identity of the
record being changed.
These dependencies connect reasoning about the current request to the
application state created by earlier requests.
The benchmark makes that connection observable across the same application.

Evaluation also depends on what generation is expected to deliver.
HumanEval and APPS test executable programs, while SWE-bench tests changes
to existing repositories~\citep{humaneval2021,apps2021,swebench2024}.
Design2Code evaluates implementation from visual
designs~\citep{design2code2025}.
WebGen-Bench and ProgramBench examine functionality after artifact
generation~\citep{webgenbench2025,programbench2026}, and AUI-Gym evaluates
task solvability on generated interfaces~\citep{auigym2025}.
\bench{} examines an application whose agent remains involved during use.
Its ordered requests test whether local successes form a complete workflow,
including revisions and return visits.

Existing agent environments provide complementary coverage of these demands.
Web tasks span compact interfaces, shopping, and realistic
websites~\citep{miniwob2017,webshop2022,mind2web2023,webarena2023,webvoyager2024}.
Mobile, desktop, enterprise, and office benchmarks assess interaction across
richer
workspaces~\citep{androidworld2024,osworld2024,omniact2024,workarena2024,officebench2024}.
AppWorld and the $\tau$ benchmarks emphasize service use and shared
state~\citep{appworld2024,taubench2024,tau2bench2025}.
AssistantBench, GAIA, and Agent's Last Exam examine broader assistant
work~\citep{assistantbench2024,gaia2023,agentslastexam2026}, while ScienceWorld
and AgentBench include stateful interaction~\citep{scienceworld2022,agentbench2023}.
Actions in these environments alter information that later steps depend on.
Alongside model-based judgments~\citep{llmjudge2023}, our evaluation retains
action evidence and saved service observations to distinguish visible
completion, persistence, and the requested result.

\section{Conclusion}
We propose Agent as App as a task for applications that adapt to a user's
changing requirements.
\liveuiagent{} connects task-specific orchestrations to a shared agent runtime,
and \bench{} evaluates continued use through ordered requests.
The raw survey records show where opening readiness and partial assertion
coverage stop short of full-workflow completion.
Submission failures, disagreements between service and displayed evidence,
and tasks left unresolved across models identify distinct challenges.
The benchmark connects these measurements to the behavior of a retained application.

\label{main-text-end}
\clearpage
\section*{AI Use Statement}
In this work, we used generative AI tools for language polishing, coding
assistance, and benchmark construction. The authors reviewed the AI-assisted
text and code and take responsibility for the final content, including the
benchmark tasks, evaluation assertions, analyses, and claims.

\section*{Ethics Statement}
Experiments use simulated AppWorld services and synthetic task identities, with operations restricted to simulated accounts. Human review examines task specifications and their expected behavior. Third-party service software and data remain subject to their upstream terms.

\section*{Reproducibility Statement}
Sections~\ref{sec:task}, \ref{sec:system}, and~\ref{sec:evaluation}
describe the task, LiveUIAgent, and model evaluation.
Appendix~\ref{app:model-survey} defines the operational metrics and survey scope.
Appendix~\ref{app:human-review} details task annotation, while
Appendix~\ref{app:case-studies} explains the recorded evidence for the main cases.
Task definitions, initial service states, generated applications, and
interaction records support verification of the reported observations.
Availability is described in Appendix~\ref{app:artifact-status}.

\begingroup\small
\bibliographystyle{iclr2027_conference}
\bibliography{revisions/yun/references-display}
\endgroup
\ifvfourappendix
\par\bigskip
\appendix
\section{Evaluation Materials}
\label{app:study-design}\label{app:artifact-status}
The model survey and case studies retain the task definitions, generated
applications, initial service context, and recorded observations needed
to interpret their outcomes.
Human annotation records connect a task specification to the corresponding
judgment and comment.
The model survey uses the original operational assertions.
The case studies follow selected executions and explain the visible and
saved evidence within each execution.

Project code carries an Apache 2.0 license.
Third-party software and data retain their upstream terms.
The complete evaluation materials are currently private.

\section{Operational Checks and Workflow Coverage}
\label{app:model-survey}\label{app:native-contract-defs}
The survey in Table~\ref{tab:multimodel} includes every assigned task,
including exposed development tasks, rejected generations, and unavailable
responses. It records one evaluation per model and uses the original
operational assertions. An opening check tests whether the declared inputs
and actions are available. Later assertions examine submitted actions,
service operations, state changes, and named targets.
The main table includes models with nonzero full-workflow rate.

For task evaluation $i$ with assigned assertions $\mathcal A_i$, the indicator
$a_{ij}\in\{0,1\}$ records whether assertion $j$ passes. The reported metrics are
\begin{equation}
\mathrm{FWR}=\frac{1}{N}\sum_i\prod_{j\in\mathcal A_i}a_{ij},\qquad
\mathrm{APR}=\frac{\sum_i\sum_{j\in\mathcal A_i}a_{ij}}{\sum_i|\mathcal A_i|}.
\label{eq:contracts}
\end{equation}
FWR weights task evaluations equally and requires all assigned assertions.
APR weights assertions equally. This distinction explains how broad partial
coverage can coexist with few complete workflows and why the metrics can
order models differently. The application examples separately relate the requested episode goals
to recorded interactions and saved service effects.

\section{Analysis of the Raw Survey Records}
\label{app:raw-analysis}
The analysis retains the same model selections as the model survey.
There are 17 selected runs over 500 common task identities.
The main model table displays the 15 runs with nonzero workflow completion.
All selected runs remain in the failure and coverage analysis.
The omitted table rows are Command A and GLM-5.3, each with zero completed
workflows out of 500 assigned tasks. Including these rows leaves the
solve-set union and the leading model comparisons unchanged.

\paragraph{Recount.}
A task passes when every recorded assertion in every prescribed episode passes.
Opening coverage is the conjunction of all opening assertions within that task.
Shared-world tasks contain later opening episodes, which are included.
The assertion pass rate divides all passed assertions by the assigned
assertion count. Rejected generations and episode errors retain their
failed assigned checks.
The recount reproduces the selected survey's task counts, opening counts,
assertion totals, and task-family results.

\paragraph{First failed episode.}
Episodes are traversed in their recorded order.
The earliest episode with a failed assertion supplies the task's failure stage.
A passing task enters the success category.
The resulting groups partition all assigned tasks.
For each first failed episode, a generation error takes precedence,
followed by a recorded episode error.
Among the remaining records, failed service and visible checks form either
a joint group or their respective exclusive groups.
The residual group contains preservation failures.
The groups describe recorded evidence rather than the internal reason
a model made an error.

\paragraph{Joint service and display checks.}
Episodes with both service and visible checks can be classified by the
conjunction within each check group.
There are 2,903 episodes passing both, 2,305 passing service checks but
failing visible checks, 1,351 with the opposite pattern, and 3,626 failing both.
A separate 10,606 records contain error placeholders.
These episode counts are distinct from task-level outcomes and from
the first-failure partition.

\paragraph{Coverage across models.}
Table~\ref{tab:survey-coverage} counts a task once if any selected run
completes it. It does not assume a way to choose the successful model
before observing the outcome.

\begin{table}[H]
\centering\small
\caption{Observed task coverage from the existing runs.
The union selects successful tasks after seeing the results and is not an
evaluated ensemble.}
\label{tab:survey-coverage}
\begin{tabular}{lrr}
\toprule
Selection or task family & Count & Assigned\\
\midrule
Best individual run & 98 & 500\\
Best pair union & 118 & 500\\
All selected runs, union & 161 & 500\\
Unsolved by every run & 339 & 500\\
\midrule
Single application, union & 139 & 302\\
Multiple tools, union & 12 & 130\\
Shared world, union & 10 & 68\\
\bottomrule
\end{tabular}
\end{table}

\subsection{Shift-alarm example}
\label{app:native-shift-case}
The user creates early-shift and late-shift alarm configurations,
switches the active configuration, returns to the application, and
switches back.
In the sol execution, creation and update API checks and the corresponding
database-change checks pass.
The revised visible result fails the check for the active late shift.
After reopening, preservation checks for both configuration labels pass.
The follow-up again passes the recorded service checks but fails the
check identifying the active early shift.
The saved configurations therefore do not establish that the interface
correctly reports the active choice.
These native service checks concern the recorded calls and changes.
They are not independent proof that every field of the alarm configuration
has the requested meaning.

\section{Human Annotation of Task Specifications}
\label{app:human-review}

\paragraph{Review criteria.}
Five human annotators inspect selected tasks' intentions, expected behavior,
interaction sequence, and prepared test conditions. They assess whether the
intention is clear and reasonable and whether the expected behavior and its
checks follow that intention.
They select one of three verdicts, reasonable, problematic, or uncertain,
and provide a comment for a problematic or uncertain verdict.
These judgments evaluate task specifications.

\paragraph{Aligning goals and tests.}
The comments expose two ways that an apparently reasonable test can fail
to express its goal. A question tracker preserved question text without
checking the answer status that the user wanted to monitor.
The task now includes a follow-up that checks that status.
An anniversary countdown left open whether a date represented a one-time
event or a recurring annual event. The goal specifies the one-time
interpretation required by its expected behavior.
The first change strengthens coverage of the promised outcome.
The second removes ambiguity about what outcome is correct.

Adjudication also checks whether an objection introduces a requirement
beyond the user's goal. Details supplied during the scripted interaction,
or quantities that follow directly from those inputs, need not appear
in the initial intention. Adjudication reconciles these comments with the task inputs and expected outcomes.

\paragraph{Annotation interface and recorded decisions.}
The review interface presents the task's initial intention and expected
behavior, together with its service dependencies and benchmark partition.
The detailed view includes the interaction episodes, prepared starting
records, assertions, and existing comments.
Annotators can therefore compare an objection with the complete task
sequence rather than only its opening requirement.

Each submitted review associates a task with a reviewer identity, a judgment,
a comment, and a timestamp.
Reasonable, problematic, and uncertain are the displayed meanings of the
stored review labels.
Problematic and uncertain submissions require a comment.
An additional free-form comment mechanism supports discussion of a task.
When a reviewer revisits the same task, the later decision remains traceable
to the earlier record.

\paragraph{Scope of the annotation.}
These records establish human review of task specifications.
They do not supply human judgments of generated-application success.
Comments provide evidence for task curation.
Application outcomes are interpreted through recorded execution and service evidence.

\section{Case Studies}
\label{app:case-studies}\label{sec:case-study}\label{app:additional-results}
The following applications relate these distinctions to a requested
revision and continued use.

\begin{figure}[!ht]
\centering
\begin{minipage}[t]{0.265\linewidth}\vspace{0pt}
\includegraphics[width=\linewidth]{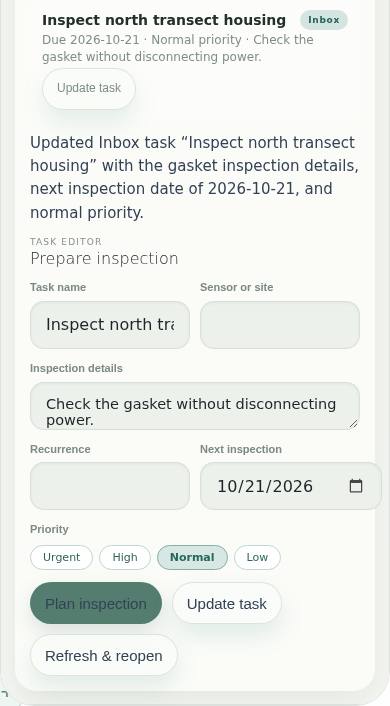}
\end{minipage}\hfill
\begin{minipage}[t]{0.70\linewidth}\vspace{0pt}
\small
\textbf{One revision, conflicting evidence}\par\medskip
\begingroup
\setlength{\fboxsep}{7pt}
\colorbox{blue!5}{\parbox{\dimexpr\linewidth-2\fboxsep\relax}{
\textbf{User request}\hfill \textbf{Interface response}\\[3pt]
Priority medium\hfill Updated, Normal selected
}}\par
\endgroup
\medskip
\begin{minipage}[t]{0.54\linewidth}
\textbf{Independent checks of the same saved record}\par\smallskip
\renewcommand{\arraystretch}{1.25}
\begin{tabular}{@{}lll@{}}
\toprule
Stage & Required & Saved \\
\midrule
After revision & medium & \textbf{low} \\
After refresh & medium & \textbf{low} \\
After reopening & medium & \textbf{low} \\
\bottomrule
\end{tabular}
\end{minipage}\hfill
\begin{minipage}[t]{0.41\linewidth}
\textbf{Evaluation outcomes}\par\smallskip
\renewcommand{\arraystretch}{1.25}
\begin{tabular}{@{}ll@{}}
\toprule
Visible completion & Pass \\
Cold preservation & Pass \\
Service compliance & \textbf{Fail} \\
\bottomrule
\end{tabular}
\end{minipage}\par\medskip
\begingroup
\setlength{\fboxsep}{7pt}
\colorbox{orange!9}{\parbox{\dimexpr\linewidth-2\fboxsep\relax}{
\textbf{The error persists on return.}\\[3pt]
The application recovers the same record, but its priority remains low at all three
independent checks.
}}
\endgroup
\end{minipage}
\caption{Visible completion can conceal an incorrect saved value.
The cropped interface detail and independent checks come from the same
wetland-sensor execution. Pixels within the rectangular crop are unchanged.
Normal is the interface label. Medium is the requested service value.
The execution also violates the permitted operation count.}
\label{fig:state-disagreement}
\end{figure}

Figure~\ref{fig:state-disagreement} follows the wetland-sensor task.
The user requests medium priority, while the interface reports a normal-priority
update. Independent reads instead find low priority after revision, refresh,
and reopening.
The visible judgment and preservation check pass.
The same record survives, but its priority fails the requested revision.

The preserved record is the intended record, but its priority remains wrong.

\paragraph{Successive requests extend the same gift queue.}
Figure~\ref{fig:giftqueue-evolution} follows an Amazon gift queue.
The user first stages a wrapped gift for Dad, then adds a wrapped gift for Ava.
The interface grows to include the additional gift while retaining the
earlier entry. The resulting queue remains visible after reopening.
The screenshots establish this visible evolution of the queue.
Service outcomes require separate evidence.

\begin{figure}[H]
\centering
\begin{minipage}[t]{0.30\linewidth}\vspace{0pt}
\small\textbf{First request}\par
One wrapped gift\par\smallskip
\includegraphics[width=\linewidth]{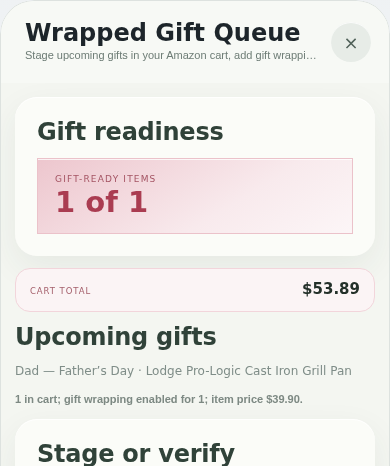}
\end{minipage}\hfill
\begin{minipage}[t]{0.30\linewidth}\vspace{0pt}
\small\textbf{Later request}\par
A second wrapped gift\par\smallskip
\includegraphics[width=\linewidth]{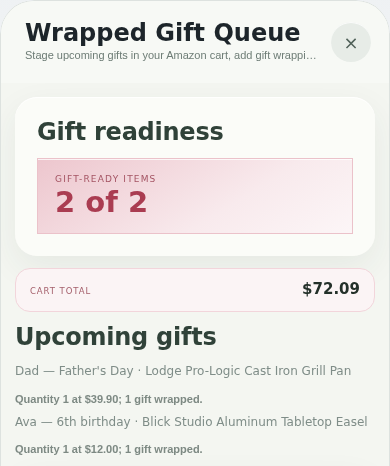}
\end{minipage}\hfill
\begin{minipage}[t]{0.30\linewidth}\vspace{0pt}
\small\textbf{On return}\par
Both entries remain\par\smallskip
\includegraphics[width=\linewidth]{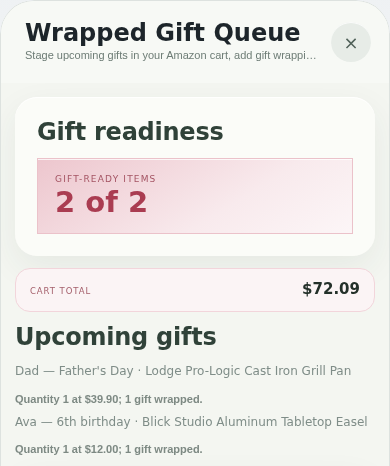}
\end{minipage}
\caption{Successive requests through the same application.
The user stages a wrapped gift for Dad and then adds one for Ava.
Cropped details show one gift entry, two entries, and the retained queue
after reopening. Pixels within each crop are unchanged.
Complete captures appear in Appendix~\ref{app:interaction-trajectories}.
Contract outcomes are evaluated separately.}
\label{fig:giftqueue-evolution}
\end{figure}

The later request extends the existing queue, and the return view retains
the earlier entries. The visible continuity illustrates continued use.
As the wetland example shows, the correctness of saved changes requires
separate evidence.

\subsection{Supporting Record Checks}
\subsubsection{Wetland priority revision}
\label{app:wetland-case}
Figure~\ref{fig:state-disagreement} combines an actual interface crop with
independent checks from the same wetland-sensor execution.
The user requests medium priority.
The interface acknowledges an update and presents Normal as the selected
label. Independent reads of the same saved record instead find low priority
after revision, refresh, and reopening.
The requested service value is medium, so the saved value does not satisfy
the revision.

The visible judgment and preservation check address different observations.
The interface provides evidence of an acknowledged interaction.
The return visit provides evidence that the application recovers its state.
The service reads establish that the recovered value remains incorrect.
The execution also exceeds the permitted mutation count.
Its failure therefore has recorded service evidence beyond the wording
of the acknowledgment.

\subsubsection{Gift queue evolution}
\label{app:giftqueue-case}
Figure~\ref{fig:giftqueue-evolution} presents unretouched crops from the
gift-queue application.
The initial request stages a wrapped gift for Dad.
The later request adds a wrapped gift for Ava, with the earlier entry
remaining visible.
A separately recorded return view retains the expanded queue.

Appendix~\ref{app:interaction-trajectories} reproduces the complete captures.
The return capture is pixel-identical to the preceding revised view.
These observations support an account of the displayed queue and its
retention. They do not, by themselves, establish every requested service
effect. The main analysis uses the example to explain how later requests
extend an existing application.

\section{Limitations}
\label{sec:limitations}
The evaluated models cover part of the range of systems that could implement
Agent as App. Future evaluation can include additional model families,
different model scales, and models specialized for interface understanding
or service use. Such coverage would help establish whether the observed
relationship between partial assertion coverage and complete workflows
persists across a broader set of capabilities.
It would also allow comparison of how different models interpret later
requests, use information established earlier, and maintain the user's
intention throughout an application session.

Generalization remains to be examined across additional tasks and service
environments. New domains may require different combinations of user input,
service operations, and retained information.
Longer interaction histories can also introduce repeated revisions and
changes in the user's preferences that are less prominent in short workflows.
Evaluation in these settings would test whether a personalized application
continues to support a changing goal as its accumulated context grows.
Extending task diversity, service coverage, and duration of use would
provide a broader account of the generality of Agent as App.

\section{LiveUIAgent Workspace}
\label{app:guios-workspace}
Figure~\ref{fig:liveuiagent-workspace} presents the material workspace
with selected applications for shopping, group messages, reading, and
meeting follow-ups.
Each application remains visible as a preview of its current recorded
content. The request area below the workspace accepts a new application
requirement or a change to existing work.

\begin{figure}[H]
\centering
\includegraphics[width=\linewidth]{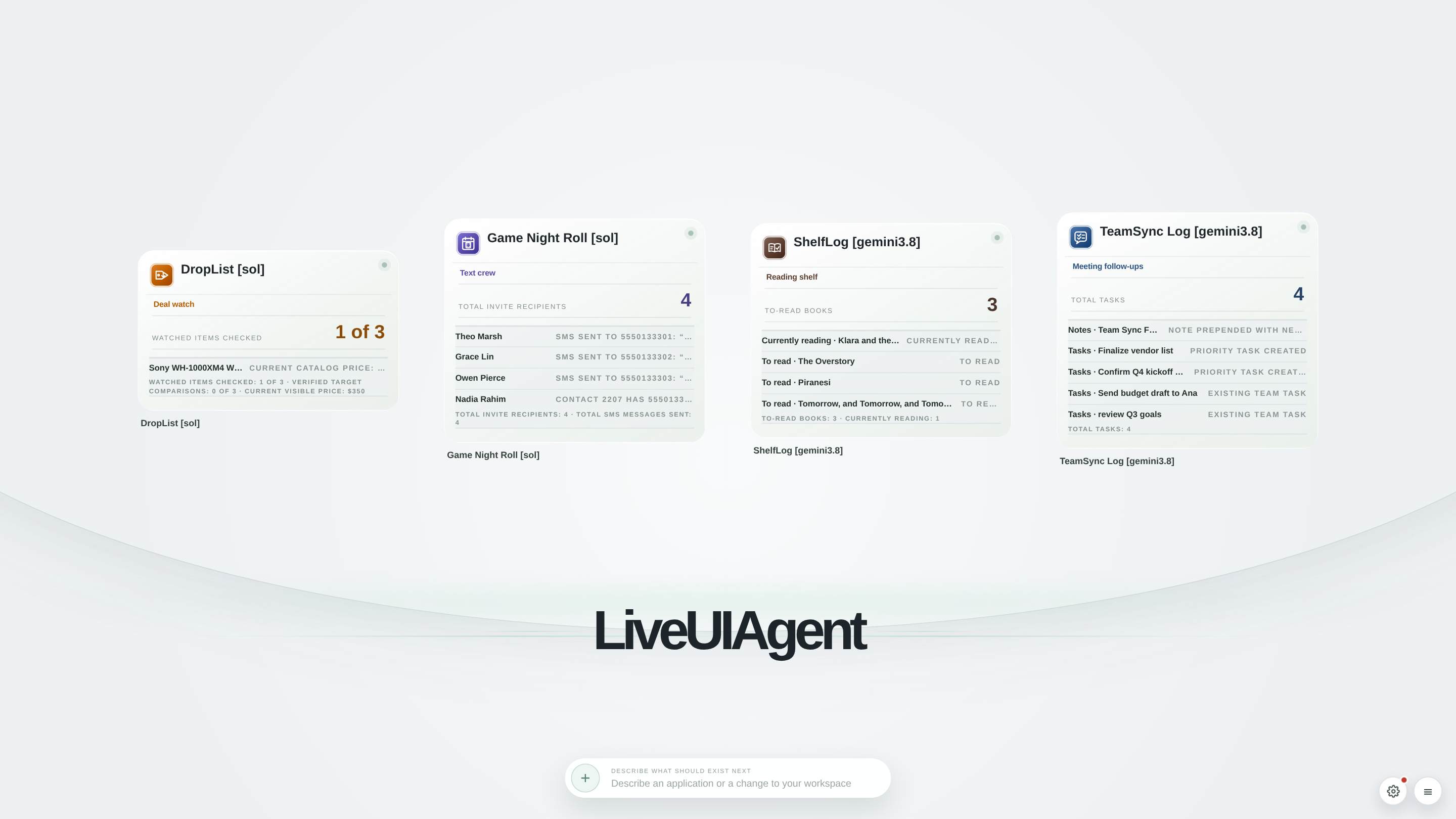}
\caption{The \liveuiagent{} material workspace with selected applications.
The previews display recorded application content, and the request area
supports continued interaction with the workspace.
The view illustrates the interface rather than a new experimental result.}
\label{fig:liveuiagent-workspace}
\end{figure}

\section{Interaction Trajectories}
\label{app:interaction-trajectories}
The sequences below follow changes to the same application across successive requests. Each panel reproduces a complete recorded interface. The Amazon sequence begins before use. The remaining panels follow the labeled task actions.
\medskip
\begin{figure}[H]
\centering
{\large\textbf{Amazon}\quad Expanding a gift queue}\par\medskip
\begin{minipage}[t]{0.48\linewidth}\vspace{0pt}\centering
{\small\textbf{1. Before use}}\par
\begin{minipage}[t][3.3\baselineskip][t]{\linewidth}\small The gift fields are empty and no operation has run.\end{minipage}\par\smallskip
\includegraphics[width=\linewidth,height=0.69\textheight,keepaspectratio]{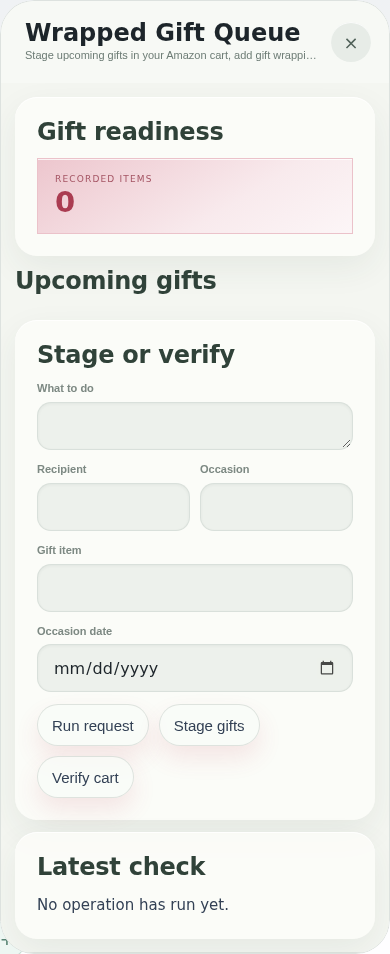}
\end{minipage}\hfill
\begin{minipage}[t]{0.48\linewidth}\vspace{0pt}\centering
{\small\textbf{2. First request}}\par
\begin{minipage}[t][3.3\baselineskip][t]{\linewidth}\small Stage the grill pan for Dad's Father's Day gift and add gift wrapping.\end{minipage}\par\smallskip
\includegraphics[width=\linewidth,height=0.69\textheight,keepaspectratio]{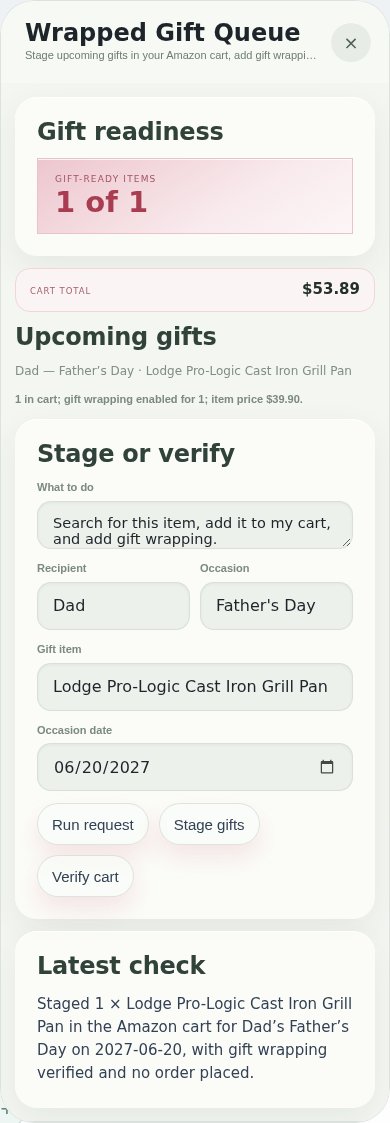}
\end{minipage}
\caption{\textbf{Amazon gift queue.} The complete initial interface is followed by the actual post-request capture, which shows one wrapped gift for Dad. These panels describe the recorded interface. Contract outcomes are evaluated separately.}
\label{fig:trajectory-amazon}
\end{figure}
\clearpage
\addtocounter{figure}{-1}
\begin{figure}[H]
\centering
{\large\textbf{Amazon}\quad Expanding a gift queue}\par\medskip
\begin{minipage}[t]{0.48\linewidth}\vspace{0pt}\centering
{\small\textbf{3. Later request}}\par
\begin{minipage}[t][3.3\baselineskip][t]{\linewidth}\small Add the tabletop easel for Ava's sixth birthday, also with gift wrapping.\end{minipage}\par\smallskip
\includegraphics[width=\linewidth,height=0.69\textheight,keepaspectratio]{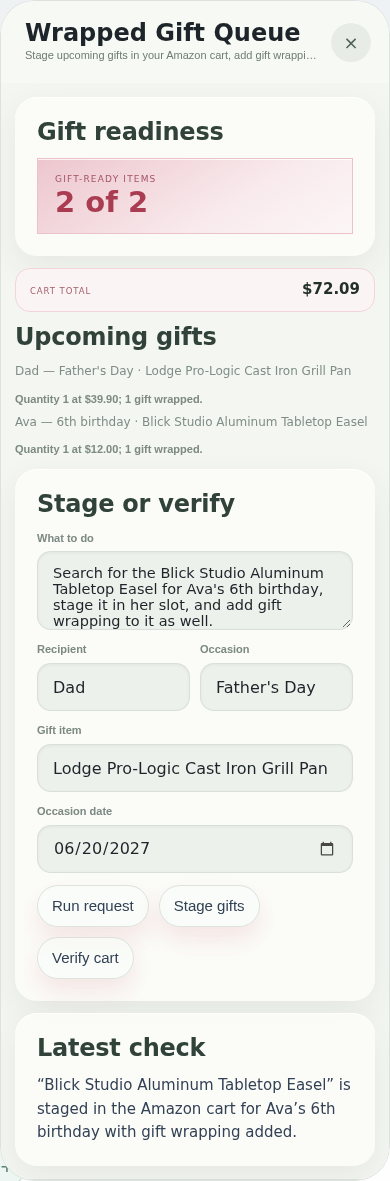}
\end{minipage}\hfill
\begin{minipage}[t]{0.48\linewidth}\vspace{0pt}\centering
{\small\textbf{4. Reopen}}\par
\begin{minipage}[t][3.3\baselineskip][t]{\linewidth}\small Close and reopen the application, preserving its current saved state.\end{minipage}\par\smallskip
\includegraphics[width=\linewidth,height=0.69\textheight,keepaspectratio]{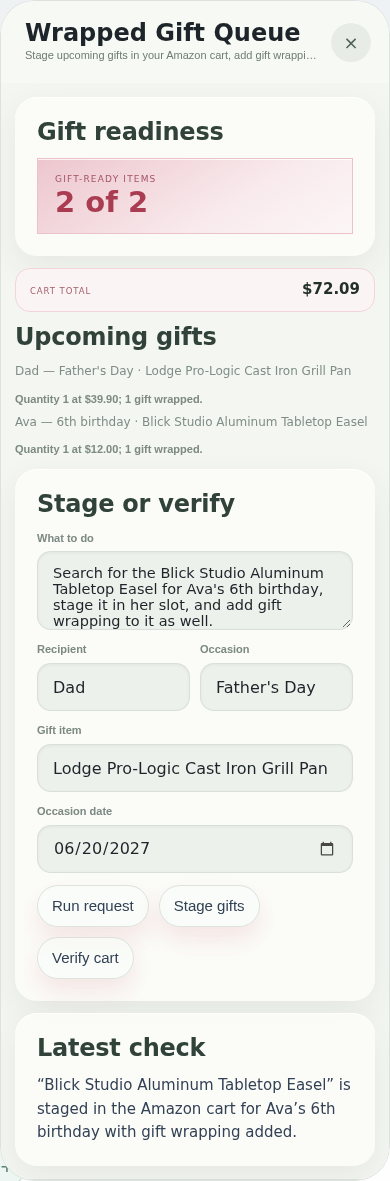}
\end{minipage}
\caption{\textbf{Amazon gift queue.} \textit{Continued.} The queue now shows wrapped gifts for Dad and Ava. The separate recorded reopening capture is pixel-identical to the post-request capture. Complete original screenshots are reproduced without retouching.}
\end{figure}

\section{LiveUIBench Case Gallery}
\label{app:liveuibench-cases}
The recorded applications below show the range of tasks in \bench{}.
Each view is an original application capture from a task execution.
The descriptions identify the request sequence and the evidence that
would be needed to judge continued use.

\begin{figure}[H]
\centering
\begin{minipage}[t]{0.48\linewidth}\centering
\textbf{Phone reminders}\par\smallskip
\includegraphics[width=\linewidth,height=0.31\textheight,keepaspectratio]{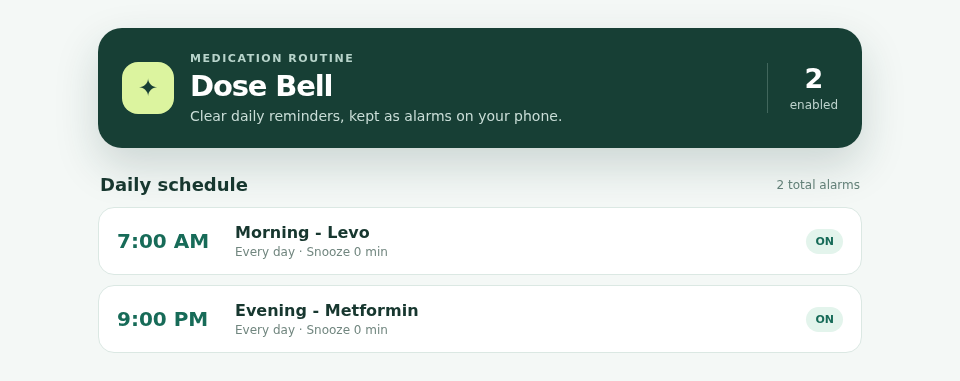}
\end{minipage}\hfill
\begin{minipage}[t]{0.48\linewidth}\centering
\textbf{Camping expenses}\par\smallskip
\includegraphics[width=\linewidth,height=0.31\textheight,keepaspectratio]{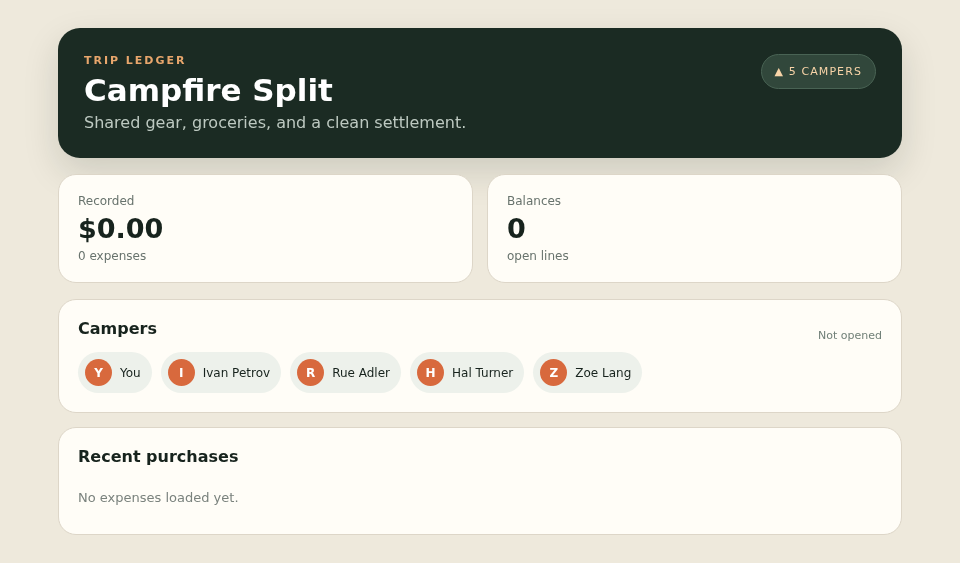}
\end{minipage}
\caption{Complete generated application views from two \bench{} cases.
The phone case combines recurring reminders with a later time change.
The expenses case combines a group, a shared expense, and a later revision.
The corresponding interaction records determine whether the displayed
state and saved service state agree.}
\label{fig:liveuibench-apps}
\end{figure}

\paragraph{Phone reminders.}
The task begins with daily reminders and later changes one reminder.
A return visit should show both reminders, with the revised time attached
to the original reminder rather than creating a second entry.
The case therefore exercises creation, revision, and retained state in one
application. The full interaction record separates the requested time
from the value displayed after reopening.

\paragraph{Camping expenses.}
The task asks for a camping group and a shared expense, followed by a gear
rental and a revision to its division among participants.
The recorded overview lists participants but shows no expense entries.
That view alone does not establish whether the requested group and expense
operations reached the service.
A later read of the service records is needed to distinguish information
displayed in the application from a saved expense.

\begin{figure}[H]
\centering
\begin{minipage}[t]{0.48\linewidth}\centering
\textbf{Todoist workspace}\par\smallskip
\includegraphics[width=\linewidth,height=0.31\textheight,keepaspectratio]{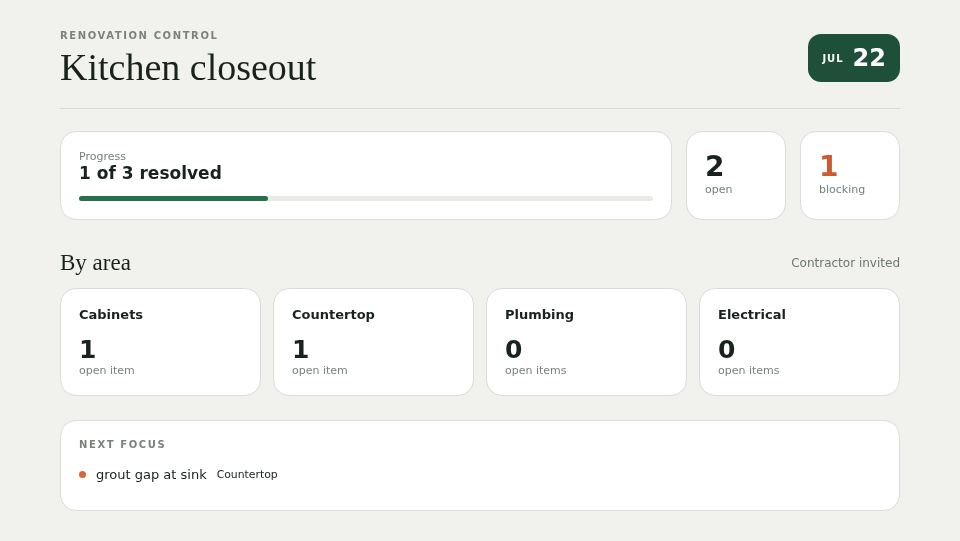}
\end{minipage}\hfill
\begin{minipage}[t]{0.48\linewidth}\centering
\textbf{File organization}\par\smallskip
\includegraphics[width=\linewidth,height=0.31\textheight,keepaspectratio]{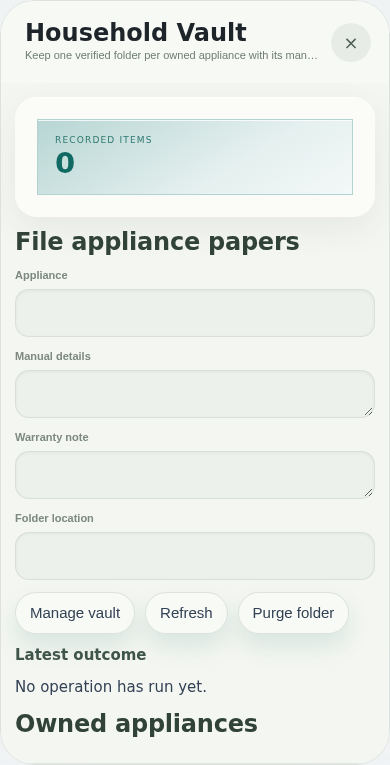}
\end{minipage}
\caption{Two additional complete application views.
The task records for these cases trace the requested item or file through
the visible interaction and the saved service state.}
\label{fig:liveuibench-more-apps}
\end{figure}

The gallery complements the detailed Amazon trajectory and the wetland
revision in Appendix~\ref{app:case-studies}. It is not a second aggregate evaluation.
Its purpose is to make the task unit visible through a generated application,
a changing request, and evidence that follows the same information across
continued use.

\fi
\end{document}